\documentclass[sigconf]{acmart}

\usepackage{macros}
\usepackage{paralist}
\usepackage{multirow}

\usepackage{listings}
\usepackage{color}

\definecolor{dkgreen}{rgb}{0,0.6,0}
\definecolor{gray}{rgb}{0.5,0.5,0.5}
\definecolor{mauve}{rgb}{0.58,0,0.82}

\AtBeginDocument{%
  \providecommand\BibTeX{{%
    \normalfont B\kern-0.5em{\scshape i\kern-0.25em b}\kern-0.8em\TeX}}}

\acmConference[]{\ }{\ }{\ }

\setcopyright{none}
\renewcommand\footnotetextcopyrightpermission[1]{} 
\begin{document}

\title{A Semantically Enriched Dataset based on Biomedical NER
\\ for the COVID19 Open Research Dataset Challenge }


\author{Hermann Kroll}
\orcid{1234-5678-9012}
\affiliation{%
  \institution{Institute for Information Systems \\ TU Braunschweig}
  \streetaddress{Mühlenpfordtstr. 23 }
  \city{Braunschweig} 
  \state{Germany} 
  \postcode{38106}
}
\email{kroll@ifis.cs.tu-bs.de}

\author{Jan Pirklbauer}
\orcid{1234-5678-9012}
\affiliation{%
  \institution{Institute for Information Systems \\ TU Braunschweig}
  \streetaddress{Mühlenpfordtstr. 23 }
  \city{Braunschweig} 
  \state{Germany} 
  \postcode{38106}
}
\email{j.pirklbauer@tu-bs.de}

\author{Johannes Ruthmann}
\orcid{1234-5678-9012}
\affiliation{%
  \institution{Institute for Information Systems \\ TU Braunschweig}
  \streetaddress{Mühlenpfordtstr. 23 }
  \city{Braunschweig} 
  \state{Germany} 
  \postcode{38106}
}
\email{j.ruthmann@tu-bs.de}

\author{Wolf-Tilo Balke}
\orcid{1234-5678-9012}
\affiliation{%
  \institution{Institute for Information Systems \\ TU Braunschweig}
  \streetaddress{Mühlenpfordtstr. 23 }
  \city{Braunschweig} 
  \state{Germany} 
  \postcode{38106}
}
\email{balke@ifis.cs.tu-bs.de}

\renewcommand{\shortauthors}{Kroll, Pirklbauer, Ruthmann, and Balke}

\begin{abstract}
Research into COVID-19 is a big challenge and highly relevant at the moment.
New tools are required to assist medical experts in their research with relevant and valuable information.
The COVID-19 Open Research Dataset Challenge (CORD-19) is a "call to action" for computer scientists to develop these innovative tools.
Many of these applications are empowered by entity information, \ie knowing which entities are used within a sentence. 
For this paper, we have developed a pipeline upon the latest Named Entity Recognition tools for Chemicals, Diseases, Genes and Species.
We apply our pipeline to the COVID-19 research challenge and share the resulting entity mentions with the community.
\end{abstract}


\keywords{Named Entity Recognition, COVID19 Research Challenge}


\maketitle

\section{Introduction}
PubMed, the most extensive library for biomedical research, contains nearly 30 million publications. 
The Allen Institute for AI selects nearly 57,000 documents as relevant for COVID19 research (V9), and around 47,000 full texts are included within this selection.
Accessing such an extensive document collection and finding relevant information is a hard task for medical researchers.
Especially in times, when results are published within a few days, keeping an overview of the latest research can be exhausting.
Novel tools are urgently needed to assist medical researchers in their workflows: novel search engines find relevant information precisely, and new access paths like summarization techniques offer new opportunities to engage the flood of information.  
These tools are typically empowered by utilizing additional side information like knowledge graphs ~\cite{dietz2018kgforir}.

Knowledge graphs are structured storages providing fact-style knowledge about entities, \eg \textit{Simvastatin is used in treatment of hypercholesterolemia}.  
In the biomedical domain, entities of interest are mainly \textit{Chemicals}, \textit{Diseases}, \textit{Genes} and \textit{Species}. 
The central problem of utilizing structured information for text retrieval is to detect, which entities are mentioned in the text.
This problem is engaged by applying a Named Entity Recognition (NER), \ie detecting important entities of in arbitrary texts.
NER tools like Spotlight (DBpedia) and WAT (Wikidata) are developed to recognize a variety of different entities in several domains~\cite{mendes_spotlight_2011,piccinno_wat_2014}.  
Unfortunately, the biomedical domain contains a variety of different entities.
Dictionary-based recognition tools might fail here because the exact entity mention within a sentence depends on the context. 
Hence, homonyms must be resolved, \eg the gene name \textit{CYP3A4} has different ids depending if the sentence talks about mouses or humans.
Yet, Named Entity Recognition tools suitable for the biomedical domain have been designed and built by experts already. 

In this paper, we utilize two biomedical NER tools, namely TaggerOne \cite{leaman_taggerone_2016} and GNormPlus \cite{wei_gnormplus_2015}, and build a pipeline to annotate arbitrary biomedical texts. 
Finally, we apply our pipeline to the COVID19 dataset.
The detected entity mentions are published in our GitHub\footnote{https://github.com/HermannKroll/CORD19BiomedicalNERDataset} repository for free reuse. 
The code will be published under the MIT license\footnote{https://opensource.org/licenses/MIT}. The data is published for free reuse under the Creative Commons Attribution 4.0 International license (CC BY 4.0)\footnote{https://creativecommons.org/licenses/by/4.0/}.
We hope that this additional entity information can serve as a solid and high-quality platform for novel tools and thus enable more research about COVID19.

\section{A Biomedical NER Pipeline}
First we will introduce a pipeline for biomedical Named Entity Recognition in arbitrary texts. 
The task of a Named Entity Recognition is to detect entity mentions in texts.
An \textbf{entity} represents a thing of interest in a specific domain, \eg Chemicals and Diseases are of interest in the biomedical domain.
Further, an entity consists of a unique id and an entity type, \eg (\textit{Simvastatin}, \textit{Chemical}) is a valid entity.  
Entities are described by a predefined vocabulary, which is typically build by experts. 
Entities might be mentioned within a written text.
Therefore, we understand text as a sequence of sentences and sentences as a sequence of tokens (single words). 
A sequence of tokens within an sentence might represent an entity. 
We call this representation an \textbf{entity mention}.
Hence, entity mentions consist of an entity and a sequence of corresponding tokens within a sentence.

The U.S. library of medicine\footnote{https://www.nlm.nih.gov} provides several expert-built tools come with a high quality for detecting entity mentions in text.
These tools can be used via command line interfaces and a freely available.
We build a pipeline upon these provided tools to automatically detect the following entity types in text: 1. Chemicals, 2. Diseases, 3. Genes and 4. Species.
Chemicals are described by the Medical Subject Heading (MeSH) vocabulary\footnote{\url{https://www.nlm.nih.gov/mesh/meshhome.html}}.
Diseases are either by MeSH terms or by OMIM\footnote{https://www.ncbi.nlm.nih.gov/omim}.
The NCBI Gene Vocabulary \footnote{https://www.ncbi.nlm.nih.gov/gene/} is utilized for the Genes' NER and the NCBI Species Taxonomy \footnote{https://www.ncbi.nlm.nih.gov/taxonomy} likewise for the Species' NER. 

Chemicals and Diseases are detected by TaggerOne \cite{leaman_taggerone_2016}, which uses a semi-Markov structured linear classifier to run named entity recognition (NER) and normalization simultaneously, thus improving performance compared to other taggers.
GNormPlus \cite{wei_gnormplus_2015} is used for detecting Genes and Species, which runs NER and normalization as two separate steps.
Both NER tools have been evaluated on real-world text corpora to determine the quality of their detected entity mentions.
Benchmarks for the relevant corpora can be found in Tables \ref{tab:taggerone} for TaggerOne and \ref{tab:gnormplus} for GNormPlus. 
NCBI Disease corpus is a testset for analysing diseases and the BioCreativeV corpus is a challenge for detecting Chemicals as well as Diseases. 
The GNormPlus evaluation is done for a Gene Normalisation testset for humans. 
Besides, GNormPlus is capable of detecting gene families in texts. 
For more details about both applications, see \cite{leaman_taggerone_2016} for TaggerOne and \cite{wei_gnormplus_2015} for GNormPlus. 

\begin{table}[]
    \centering
    \caption{Benchmark results of TaggerOne \cite{leaman_taggerone_2016}}
   \begin{tabular}{|c|c|c|c|}
    \hline
    \textbf{Corpus} & \textbf{Precision} & \textbf{Recall} & \textbf{F-measure}\\
    \hline    
    NCBI Disease  & 81.5\% & 80.8\% & 82.9\% \\
    \hline
    BioCreativeV CD-R & 94.2\% & 88.8\% & 91.4\% \\
    \hline
    \end{tabular}
    \label{tab:taggerone}
\end{table}

\begin{table}[]
    \centering
    \caption{Benchmark results of GNormPlus (Human) \cite{wei_gnormplus_2015}}
    \begin{tabular}{|c|c|c|c|}
    \hline
    \textbf{Corpus} & \textbf{Precision} & \textbf{Recall} & \textbf{F-measure}\\
    \hline    
    BioCreative II GN & 87.1\% & 86.4\% & 86.7\% \\         
    \hline
    \end{tabular}
    \label{tab:gnormplus}
\end{table}

\paragraph{Pipeline}
We have developed a pipeline utilizing TaggerOne and GNormPlus for biomedical NER. 
Our pipeline expects texts in a so-called PubTator format, see \cite{wei_pubtator_2013} and the description on\footnote{\url{https://www.ncbi.nlm.nih.gov/research/pubtator/}}.
As an input, the pipeline supports 1. a single PubTator file, 2. a composed PubTator file and 3. a directory of PubTator files. 
A composed PubTator file consists of the content of two PubTator files separated by two newlines.
Besides, we support the tagging of multiple files in parallel. 
Therefore, we implemented a splitting of the input and parallel working of the underlying tools.
The recognition steps stores it's produced data in a relational database.
Finally, the pipeline exports the annotated entity mentions in a desired format like PubTator or JSON.

\section{The COVID-19 Open Research Dataset}
\begin{table}[]
    \centering
    \caption{Document Counts of CORD19 Sources}    
    \begin{tabular}{|r|l|}
        \hline
        \textbf{General} & \\
        \hline
        Number of Documents & 57.4K \\
        Number of full texts & 43.5K \\
        \hline
        \textbf{JSON parses by source} & \\
        \hline
        PubMedCentral (PMC) & 49.7K \\
        Elsevier & 24.8K \\
        medRxiv & 2.3K \\
        ArXiv & 1.2K \\
        bioRxiv & 1.1K \\
        Chan Zuckerberg Initiative (CZI) & 0.2K \\
        \hline
    \end{tabular}
    \label{tab:cord19}
\end{table}

Research into COVID-19 is a big challenge and highly relevant at the moment.
Therefore, scientists in the medical field must be assisted by innovative tools to access the current state of literature efficiently.
The COVID-19 Open Research Dataset Challenge (CORD-19) \cite{aiai_cord19_2020} is a "call to action" for computer scientists in the natural language processing (NLP) and data mining field to develop such innovative tools.
The dataset in version 9 consists of ca. 57,000 scholarly articles, of which ca. 44,000 have a PDF parse of their full text attached to them. 
Articles are taken from various sources, most prominently the PubMedCentral collection. 
The document statistics of the dataset in version 9 can be seen in Table \ref{tab:cord19}.
Some documents are accessible in multiple sources and are counted more than once in the statistics.
The abstracts and full texts of the documents are given paragraph wise in a JSON-Format, so the texts can easily be extracted and processed. 
Entity-centric information access plays a key role in the medical domain \cite{herskovic2007pubmedqueryanalysis}. 
Hence, we run our pipeline upon the challenge dataset  to assist the community with valuable entity information.

\begin{table}[]
    \centering
    \caption{Number of Detected Entity Mentions for the CORD-19 (Abstracts and Fulltexts)}
    \begin{tabular}{|r|c|c|c|c|}
        \hline
        \textbf{Corpus} & Chemicals & Diseases & Genes & Species  \\
        \hline
        Abstracts & 99K & 145K  & 59K  &  165K \\
        \hline
        Fulltexts  & 3,407K & 4,039K  & 2,232K  &  4,667K \\
        \hline
    \end{tabular}
      \label{tab:cord19entitymentions}
\end{table}

\subsection{Detected Entity Mentions}
We report the number of the resulting entity mentions for each entity type.
We create two different dumps: one dump contains entity mentions within titles and abstracts and the second dump contains entity mentions in the title, abstract and fulltexts of the documents. 
Table~\ref{tab:cord19entitymentions} lists the number of entity mention for both dumps grouped by the entity types.
Our pipeline detects nearly 99K Chemicals, 145K Diseases, 59K Genes and 165K Species in titles and abstracts. 
For fulltexts, the pipeline detects around 3.4M Chemicals, 4.0M Diseases, 2.2M Genes and 4,7M Species.
We estimate the annotation's quality to be comparable to the reported quality in the tools' original publications. 

\subsection{Dump of the Entity Mentions}
We publish the obtained entity mentions as two JSON files. 
The first file contains the entity mentions for titles and abstracts.
The second file contains the entity mentions for titles, abstracts as well as fulltexts.
We process the CORD19 fulltexts by selecting the available JSON files. 
These JSON files contain fulltexts as sequences of body texts. 
Hence, a fulltext document consists of a title, an abstract and a sequence of body texts.
We publish the corresponding entity mentions suitable for the given structure.
Therefore, each entity mentions contains an entity location in texts including:
\begin{enumerate}
    \item a paragraph representing the position in the text. 0 is an entity mention in the title, 1 is an entity mention in the abstract and 2 is an entity mention in the first body text field and so on.
    \item a start position representing the position of the first entity's character within the corresponding text (title, abstract, body text element). 
    \item an end position representing the position of the last entity's character within the corresponding text.
\end{enumerate}
As an example, an entity location with paragraph 5, start  5 and end 10 means that the entity is mentioned in the third body text field starting at character position 5 and ending at character position 10. 
The first character has the position 0.
An entity mention contains the following components:
\begin{enumerate}
    \item an entity location,
    \item an entity string representing the entity's token sequence in the text,
    \item an entity type (Chemical, Disease, Gene and Species), and
    \item an entity id corresponding to the previously described vocabularies. 
\end{enumerate}

The computed entity mentions are shared within a JSON file. 
The JSON file consists of a dictionary, where each CORD19 document id is mapped to a list of entity mentions.
A short prototypical snapshot of the exported JSON file is shown below:
\lstset{language=Python}
\begin{lstlisting}
[
  <paper_id: str>: [ #For every JSON-parse of the dataset 
    {   # For every entity mention
      "location": {
        "paragraph": <int>  # 0 = title, 1 = abstract
                             # > 1 = body text
        "start": <int> # 0 = first character of paragraph
        "end": <int>
      },
      "entity_str": <str> # entity mention in source text
      "entity_type": <"Chemical"|"Disease"|"Gene"|"Species">
      "entity_id": <str> # e.g. MESH-Identifier
    },...
  ],...
]

\end{lstlisting}
More details can be found in our regularly updated GitHub repository.

\section{Summary and Outlook}
In this paper, we discussed the importance and usefulness of entity mentions for retrieval applications. We developed an effective pipeline to automatically annotate biomedical entity mentions in arbitrary texts. Moreover, we built our pipeline on top of the latest available biomedical NER tools to ensure the quality of our entity mentions.

Applying our pipeline to the COVID-19 open research dataset, we published the resulting entity mentions as a semantically enriched dataset for free reuse on GitHub. We will continuously update our GitHub repository whenever new versions of the COVID-19 dataset are published.

\bibliographystyle{ACM-Reference-Format}
\bibliography{references}


\begin{thebibliography}{8}


\ifx \showCODEN    \undefined \def \showCODEN     #1{\unskip}     \fi
\ifx \showDOI      \undefined \def \showDOI       #1{#1}\fi
\ifx \showISBNx    \undefined \def \showISBNx     #1{\unskip}     \fi
\ifx \showISBNxiii \undefined \def \showISBNxiii  #1{\unskip}     \fi
\ifx \showISSN     \undefined \def \showISSN      #1{\unskip}     \fi
\ifx \showLCCN     \undefined \def \showLCCN      #1{\unskip}     \fi
\ifx \shownote     \undefined \def \shownote      #1{#1}          \fi
\ifx \showarticletitle \undefined \def \showarticletitle #1{#1}   \fi
\ifx \showURL      \undefined \def \showURL       {\relax}        \fi
\providecommand\bibfield[2]{#2}
\providecommand\bibinfo[2]{#2}
\providecommand\natexlab[1]{#1}
\providecommand\showeprint[2][]{arXiv:#2}

\bibitem[\protect\citeauthoryear{Dietz, Kotov, and Meij}{Dietz
  et~al\mbox{.}}{2018}]%
        {dietz2018kgforir}
\bibfield{author}{\bibinfo{person}{Laura Dietz}, \bibinfo{person}{Alexander
  Kotov}, {and} \bibinfo{person}{Edgar Meij}.} \bibinfo{year}{2018}\natexlab{}.
\newblock \showarticletitle{Utilizing Knowledge Graphs for Text-Centric
  Information Retrieval}. In \bibinfo{booktitle}{\emph{The 41st International
  ACM SIGIR Conference on Research \& Development in Information Retrieval}}
  (Ann Arbor, MI, USA) \emph{(\bibinfo{series}{SIGIR ’18})}.
  \bibinfo{publisher}{Association for Computing Machinery},
  \bibinfo{address}{New York, NY, USA}, \bibinfo{pages}{1387–1390}.
\newblock
\showISBNx{9781450356572}
\urldef\tempurl%
\url{https://doi.org/10.1145/3209978.3210187}
\showDOI{\tempurl}


\bibitem[\protect\citeauthoryear{for AI, et~al., and House}{for AI
  et~al\mbox{.}}{2020}]%
        {aiai_cord19_2020}
\bibfield{author}{\bibinfo{person}{Alan~Institute for AI},
  \bibinfo{person}{Anthony~Goldbloom et al.}, {and} \bibinfo{person}{The~White
  House}.} \bibinfo{year}{2020}\natexlab{}.
\newblock \bibinfo{title}{COVID-19 Open Research Dataset Challenge (CORD-19),
  Version 9}.
\newblock
\newblock
\newblock
\shownote{Retrieved April 27, 2020 from
  \url{https://www.kaggle.com/dataset/08dd9ead3afd4f61ef246bfd6aee098765a19d9f6dbf514f0142965748be859b/version/9}.}


\bibitem[\protect\citeauthoryear{Herskovic, Tanaka, Hersh, and
  Bernstam}{Herskovic et~al\mbox{.}}{2007}]%
        {herskovic2007pubmedqueryanalysis}
\bibfield{author}{\bibinfo{person}{Jorge~R. Herskovic}, \bibinfo{person}{Len~Y.
  Tanaka}, \bibinfo{person}{William Hersh}, {and} \bibinfo{person}{Elmer~V.
  Bernstam}.} \bibinfo{year}{2007}\natexlab{}.
\newblock \showarticletitle{{A Day in the Life of PubMed: Analysis of a Typical
  Day's Query Log}}.
\newblock \bibinfo{journal}{\emph{Journal of the American Medical Informatics
  Association}} \bibinfo{volume}{14}, \bibinfo{number}{2} (\bibinfo{date}{03}
  \bibinfo{year}{2007}), \bibinfo{pages}{212--220}.
\newblock
\showISSN{1067-5027}


\bibitem[\protect\citeauthoryear{Leaman and Lu}{Leaman and Lu}{2016}]%
        {leaman_taggerone_2016}
\bibfield{author}{\bibinfo{person}{Robert Leaman} {and}
  \bibinfo{person}{Zhiyong Lu}.} \bibinfo{year}{2016}\natexlab{}.
\newblock \showarticletitle{{TaggerOne: joint named entity recognition and
  normalization with semi-Markov Models}}.
\newblock \bibinfo{journal}{\emph{Bioinformatics}} \bibinfo{volume}{32},
  \bibinfo{number}{18} (\bibinfo{date}{06} \bibinfo{year}{2016}),
  \bibinfo{pages}{2839--2846}.
\newblock
\showISSN{1367-4803}
\urldef\tempurl%
\url{https://doi.org/10.1093/bioinformatics/btw343}
\showDOI{\tempurl}
\showeprint{https://academic.oup.com/bioinformatics/article-pdf/32/18/2839/24406872/btw343.pdf}


\bibitem[\protect\citeauthoryear{Mendes, Jakob, Garc\'{\i}a-Silva, and
  Bizer}{Mendes et~al\mbox{.}}{2011}]%
        {mendes_spotlight_2011}
\bibfield{author}{\bibinfo{person}{Pablo~N. Mendes}, \bibinfo{person}{Max
  Jakob}, \bibinfo{person}{Andr\'{e}s Garc\'{\i}a-Silva}, {and}
  \bibinfo{person}{Christian Bizer}.} \bibinfo{year}{2011}\natexlab{}.
\newblock \showarticletitle{DBpedia Spotlight: Shedding Light on the Web of
  Documents}. In \bibinfo{booktitle}{\emph{Proceedings of the 7th Int. Conf. on
  Semantic Systems}} (Graz, Austria) \emph{(\bibinfo{series}{I-Semantics
  ’11})}. \bibinfo{publisher}{Association for Computing Machinery},
  \bibinfo{address}{New York, NY, USA}, \bibinfo{pages}{1–8}.
\newblock
\showISBNx{9781450306218}


\bibitem[\protect\citeauthoryear{Piccinno and Ferragina}{Piccinno and
  Ferragina}{2014}]%
        {piccinno_wat_2014}
\bibfield{author}{\bibinfo{person}{Francesco Piccinno} {and}
  \bibinfo{person}{Paolo Ferragina}.} \bibinfo{year}{2014}\natexlab{}.
\newblock \showarticletitle{From TagME to WAT: A New Entity Annotator}. In
  \bibinfo{booktitle}{\emph{Proceedings of the First Int. Workshop on Entity
  Recognition \& Disambiguation}} (Gold Coast, Queensland, Australia)
  \emph{(\bibinfo{series}{ERD ’14})}. \bibinfo{publisher}{Association for
  Computing Machinery}, \bibinfo{address}{New York, NY, USA},
  \bibinfo{pages}{55–62}.
\newblock
\showISBNx{9781450330237}


\bibitem[\protect\citeauthoryear{Wei, Kao, and Lu}{Wei et~al\mbox{.}}{2013}]%
        {wei_pubtator_2013}
\bibfield{author}{\bibinfo{person}{Chih-Hsuan Wei}, \bibinfo{person}{Hung-Yu
  Kao}, {and} \bibinfo{person}{Zhiyong Lu}.} \bibinfo{year}{2013}\natexlab{}.
\newblock \showarticletitle{{PubTator: a web-based text mining tool for
  assisting biocuration}}.
\newblock \bibinfo{journal}{\emph{Nucleic Acids Research}}
  \bibinfo{volume}{41}, \bibinfo{number}{W1} (\bibinfo{date}{05}
  \bibinfo{year}{2013}), \bibinfo{pages}{W518--W522}.
\newblock
\showISSN{0305-1048}
\urldef\tempurl%
\url{https://doi.org/10.1093/nar/gkt441}
\showDOI{\tempurl}
\showeprint{https://academic.oup.com/nar/article-pdf/41/W1/W518/3859973/gkt441.pdf}


\bibitem[\protect\citeauthoryear{Wei, Kao, and lu}{Wei et~al\mbox{.}}{2015}]%
        {wei_gnormplus_2015}
\bibfield{author}{\bibinfo{person}{Chih-Hsuan Wei}, \bibinfo{person}{Hung-Yu
  Kao}, {and} \bibinfo{person}{Zhiyong lu}.} \bibinfo{year}{2015}\natexlab{}.
\newblock \showarticletitle{GNormPlus: An Integrative Approach for Tagging
  Genes, Gene Families, and Protein Domains}.
\newblock \bibinfo{journal}{\emph{BioMed research international}}
  \bibinfo{volume}{2015} (\bibinfo{date}{09} \bibinfo{year}{2015}),
  \bibinfo{pages}{918710}.
\newblock
\urldef\tempurl%
\url{https://doi.org/10.1155/2015/918710}
\showDOI{\tempurl}


\end{thebibliography}


\end{document}